\documentclass[11pt]{iopart}
\usepackage{times}
\usepackage{graphicx}

\begin{document}

\title{Classical Harmonic Oscillator with Dirac-like Parameters and Possible Applications}
\author{H C Rosu\footnote{e-mail: hcr@ipicyt.edu.mx \hfill {\tiny files hokiop2 in iop,hosept.pdf/Sept28/2004}}, O Cornejo-P\'erez, R L\'opez-Sandoval}
\address{Instituto Potosino de Investigaci\'on Cient\'{\i}fica y Tecnol\'ogica, (IPICyT),\\
Apdo Postal 3-74 Tangamanga, 78231 San Luis Potos\'{\i}, MEXICO}
\begin{abstract}
We obtain a class of parametric oscillation modes that we call K-modes with damping and absorption that are connected to the classical harmonic oscillator modes
through the ``supersymmetric" one-dimensional matrix procedure similar to relationships of the same type between Dirac and Schr\"odinger equations in particle 
physics. When a single coupling parameter, denoted by K, is used, it characterizes both the damping and the dissipative features of these modes. 
Generalizations to several K parameters are also possible and lead to analytical results.
If the problem is passed to the physical optics (and/or acoustics) context by switching from the oscillator equation to the 
corresponding Helmholtz equation, one may hope to detect the K-modes as waveguide modes of specially designed waveguides and/or cavities. 
\end{abstract}

{

PACS number(s):
12.60.Jv, 11.30.Pb \hfill  arXiv number: math-ph/0402065

\bigskip

{\bf 1}. {\em Introduction.} $\quad$ \\
Factorizations of differential operators describing simple mechanical motion have been only occasionally used in the past, although in quantum mechanics 
the procedure led to a vast literature under the name of supersymmetric quantum mechanics initiated by a paper of Witten \cite{w81}. 
However, as shown by Rosu and Reyes \cite{rr98}, for the damped Newtonian free oscillator the factorization method could generate 
interesting results even in an area settled more than three centuries ago. In the following, we apply some of the supersymmetric schemes to the basic classical harmonic oscillator. In particular, we show how a known connection in particle physics between Dirac and Schr\"odinger equations could lead in the case of harmonic motion to 
chirped (i.e., time-dependent) frequency oscillator equations whose solutions are
a class of oscillatory modes depending on one more parameter, denoted by K in this work, besides the natural circular frequency $\omega _0$. The parameter K characterizes 
both the damping and the losses of these ``supersymmetric" partner modes. Moreover, we do not limit this study to one K parameter extending it to several such parameters
still getting analytic results.
Guided by mathematical equivalence, possible applications in several areas of physics are identified.

\bigskip


{\bf 2}. {\em Classical harmonic oscillator: The Riccati approach.} $\quad$ \\
The harmonic oscillator can be described by one of the simplest Riccati equation
\begin{equation} \label{ricc}
{\rm u^{'}+u^2+\kappa\omega _0^2=0, \qquad  \kappa =\pm 1~,}
\end{equation}
where the plus sign is for the normal case whereas the minus sign is for the up side down case. 
Indeed,
employing ${\rm u=\frac{w^{'}}{w}}$ one gets the
harmonic oscillator differential equation
\begin{equation} \label{schr1}
{\rm w^{''}+\kappa\omega_{0}^2w=0~,}
\end{equation}
with the solutions
$$
{{\rm w_{b}}=
\left\{ \begin{array}{ll}
{\rm W_+\cos(\omega _0 t +\varphi _{+})} & \mbox{if $\kappa =1$}\\
{\rm W_{-}{\rm sinh}(\omega _0 t +\varphi _{-})} & \mbox{if $\kappa =-1$~,}
\end{array} \right.}
$$
where ${\rm W_{\pm }}$ and $\varphi _{\pm}$
are amplitude and phase parameters, respectively, which can be ignored in the following.

The particular Riccati solution
of Eq.~(\ref{ricc}) are
$$
{\rm u_{p}=
\left\{ \begin{array}{ll}
-\omega _0{\rm tan (\omega _0 t)} & \mbox{if $\kappa =1$}\\
\omega _0{\rm coth (\omega _0 t)} & \mbox{if $\kappa =-1$~.}
\end{array} \right.}
$$
It is well known that the particular Riccati solutions enter as nonoperatorial part in
the common factorizations of the second-order linear differential
equations that are directly related to the Darboux isospectral transformations
\cite{ms}.

Thus, for Eq.~ (\ref{schr1}) one gets (${\rm D_{t}=\frac{d}{dt}}$)
\begin{equation} \label{w3}
{\rm \left(D_{t}+u_{p}\right)
\left(D_{t}-u_{p}\right)w=
w^{''}+(-u_{p}^{'}-u_{p}^{2})w=0}~.
\end{equation}
To fix the ideas, we shall use the terminology of Witten's supersymmetric quantum mechanics and call Eq.~(\ref{w3}) the 
bosonic equation. We stress here that the supersymmetric terminology is used in this paper only for convenience and should not be taken literally.
Thus, the supersymmetric partner (or fermionic) 
equation of Eq.~(\ref{w3}) is obtained by reversing the factorization brackets
\begin{equation} \label{f}
{\rm
\left(D_{t}-u_{p}\right)
\left(D_{t}+u_{p}\right)w_f=
w^{''}+(u_{p}^{'}-u_{p}^2)w={\rm w^{''}
+\omega ^2_{f}(t)w=0}}~,
\end{equation}
which is related to the fermionic Riccati equation
\begin{equation} \label{fR}
{\rm u^{'}-u^2-\omega ^2_{{\rm f}}(t)=0~,}
\end{equation}
where the free term $\omega _{\rm  f}^2$ is the following function of time
$$
{\rm \omega ^2_{f}(t)=u_{p}^{'}-u_{p}^2=
\left\{ \begin{array}{ll}
{\rm \omega _0^2(-1-2{\rm tan}^2 \omega _0 t)} & \mbox{if $\kappa =1$}\\
{\rm \omega _0^2(1-2{\rm coth}^2 \omega _0 t)} & \mbox{if $\kappa =-1$~.}
\end{array} \right.}
$$
The solutions (fermionic zero modes) of Eq.~(\ref{f}) are given by
$$
{{\rm w_{f}}=
\left\{ \begin{array}{ll}
{\rm \frac{-\omega _0}{\cos (\omega _0 t)}} & \mbox{if $\kappa =1$}\\
{\rm \frac{\omega _0}{sinh (\omega _0 t)}} & \mbox{if $\kappa =-1$~,}
\end{array} \right.}
$$
and thus present strong periodic singularities in the first case and just one singularity at the origin in the second case.
These `partner' oscillators, as well as those to be discussed in the following, are parametric oscillators, i.e., of time-dependent
frequency. Moreover, their frequencies can become infinite (periodically). In general, signals of this type are known as chirps.
`Infinite' chirps could be produced, in principle, in very special astrophysical circumstances, e.g., close to
black hole  horizons \cite{J96}. 

\bigskip

\noindent
{\bf 3}. {\em Matrix formulation.} $\quad$ \\
Using the Pauli matrices 
$
\sigma _y=
\left( \begin{array}{cc}
0 & -{\rm i }\\
{\rm i} & 0\end{array} \right )$ and  
$\sigma _x=\left( \begin{array}{cc}
0 & 1\\
1 & 0 \end{array} \right )~, 
$
we write the matrix equation 
\begin{equation} \label{HD}
\hat{\cal D}_0W\equiv {\rm  [\sigma _y D_{t}+\sigma _x (iu_p)]}W=0~,
\end{equation}
where $W=\left( \begin{array}{cc}
{\rm w_1}\\
{\rm w_2}\end{array} \right )$ is a two component spinor. 
Eq.~(\ref{HD}) is equivalent to the following decoupled equations
\begin{eqnarray}
({\rm i D_{t}+i u_p)w_1=0}\\
({\rm -i D_{t}+i u_p)w_2=0}~.
\end{eqnarray}
Solving these equations one gets ${\rm w}_1\propto \omega _0/\cos ({\rm \omega _0}t)$ and ${\rm w}_2\propto \omega _0\cos({\rm \omega _0}t)$ 
for the $\kappa =1$ case
and ${\rm w}_1\propto \omega _0/{\rm sinh} ({\rm \omega _0}t)$ and ${\rm w}_2\propto \omega _0{\rm sinh}({\rm \omega _0}t)$ 
for the $\kappa =-1$ case.
Thus, we obtain 
\begin{equation}\label{W1}
W=\left( {\rm \begin{array}{cc}
{\rm w_1}\\
{\rm w_2}\end{array}} \right )=\left( {\rm \begin{array}{cc}
{\rm w_f}\\
{\rm w_b}\end{array}} \right )~.
\end{equation}
This shows that the matrix 
equation is equivalent to the two second-order linear differential  
equations of bosonic and fermionic type, Eq.~(\ref{schr1}) and Eq.~(\ref{f}), respectively, a result quite well known in particle physics.
Indeed, a comparison with the true Dirac equation with a Lorentz scalar potential $\rm S(x)$
\begin{equation} \label{tD}
{\rm  [-i\sigma _y D_{x}+\sigma _x (m +S(x))]}W={\rm E}W~,
\end{equation}
shows that Eq.~(\ref{HD}) corresponds to a Dirac spinor of `zero mass' and `zero energy' in an {\em imaginary} scalar `potential' $\rm i u_p(t)$. 
We remind that a detailed discussion of the Dirac equation in the supersymmetric approach has been provided by Cooper {\em et al} \cite{cooper} in 1988.
They showed that the Dirac equation with a Lorentz scalar potential is associated with a susy pair of Schroedinger Hamiltonians.
This result has been used later by many authors in the particle physics context \cite{literature}.

\bigskip

\noindent
{\bf 4}. {\em Extension through parameter {\rm K}.} $\quad$\\ 
We now come to the main issue of this work. Consider the slightly more general Dirac-like equation 
\begin{equation} \label{HDM}
\hat{\cal D}_{{\rm K}}W\equiv {\rm [\sigma _y D_{t}+\sigma _x (iu_p +K)]W=KW}~,
\end{equation}
where K is a (not necessarily positive) real constant. On the left hand side of the equation, $\rm K$ stands as 
an (imaginary) mass parameter of the Dirac spinor, whereas 
on the right hand side it corresponds to the energy parameter. 
Thus, we have an equation equivalent to a Dirac equation for a spinor
of mass $\rm i K$ at the fixed energy $E={\rm i K}$. This equation  
can be written as the following system of coupled equations
\begin{eqnarray}
{\rm iD_{t}w _1+(iu_p+K)w _1=Kw _2}\\
{\rm -iD_{t}w _2+(iu_p+K)w _2=Kw _1}~.
\end{eqnarray}

The decoupling can be achieved by applying the operator in Eq.~(13) to Eq.~(12) . For the fermionic spinor 
component one gets
\begin{equation} \label{comp1} 
{\rm D^{2}_{t}w_1^{+}-\omega _0^2\Big[(1+2\tan ^2 \omega _0 t)+i\frac{2K}{\omega _0}\tan \omega _0 t\Big] w_1^{+}=0  \quad {\rm for} \,\, \kappa =1}
\end{equation}
\begin{equation} \label{comp1b} 
{\rm D^{2}_{t}w_1^{-}+\omega _0^2\Big[(1-2{\rm coth}^2 \omega _0 t)+i\frac{2K}{\omega _0}{\rm coth} \,\omega _0 t\Big] w_1^{-}=0  
\quad  {\rm for} \,\, \kappa =-1}~,
\end{equation}
whereas the bosonic component fulfills
\begin{equation} \label{comp2} 
{\rm D^{2}_{t}w_2^{+}+\omega _0^2\Big[1-i\frac{2K}{\omega _0}\tan \omega _0t\Big] w_2^{+}=0  \quad {\rm for} \,\,  \kappa =1}
\end{equation}
\begin{equation} \label{comp2b} 
{\rm D^{2}_{t}w_2^{-}-\omega _0^2\Big[1-i\frac{2K}{\omega _0}{\rm coth} \,\omega _0t\Big] w_2^{-}=0  \quad {\rm for} \,\, \kappa =-1}~.
\end{equation}

\bigskip
The solutions of the bosonic equations are expressed in terms of
the Gauss hypergeometric functions $_{2}{\rm F}_{1}$
\begin{eqnarray}
{\rm w} _{2}^{+}{\rm (t;\alpha_{+},\beta_{+})=\alpha_{+}z_{1}^{(p-\frac{1}{2})}
z_{2}^{(q-\frac{1}{2})}\,
_{2} F_{1}\left[p+q,p+q-1,2p\,;-\frac{1}{2}z_{1}\right]}\nonumber\\
 - {\rm \beta_{+}e^{-2ip\pi}4^{(p-\frac{1}{2})}z_{1}^{-(p-\frac{1}{2})} z_{2}^{(q-\frac{1}{2})}\,
_{2}F_{1}\left[q-p,q-p+1,2-2p\,;-\frac{1}{2}z_{1}\right]}
\end{eqnarray}
and
\begin{eqnarray}
{\rm w_{2}^{-}(t;\alpha_{-},\beta_{-}) = \alpha_{-}z_{3}^{r} z_{4}^{s}\,
_{2}F_{1}\left[r+s,r+s+1,1+2r;\frac{1}{2}z_{3}\right]}\nonumber\\
 + {\rm \beta_{-}4^r z_{3}^{-r} z_{4}^{s}\,
_{2}F_{1}\left[s-r+1,s-r,1-2r;\frac{1}{2}z_{3}\right]}~,
\end{eqnarray}
where the variables ${\rm z}_{i}$ ($i=1,...,4$) are given in the
following form: 
$${\rm z_{1}=i\tan(\omega _0 t)-1, \quad z_{2}=i\tan(\omega _0 t)+1,\quad
z_{3}={\rm coth}(\omega _0 t)+1, \quad z_{4}={\rm coth}(\omega _0 t)-1},
$$ 
respectively. The parameters are the following:
$$
{\rm p=\frac{1}{2}\left(1+\sqrt{1-\frac{2K}{\omega _0}}\right),\quad
q = \frac{1}{2}\left(1+\sqrt{1+\frac{2K}{\omega _0}}\right), \quad
r = \frac{1}{2}\sqrt{1+i\frac{2K}{\omega _0}}, \quad
s = \frac{1}{2}\sqrt{1-i\frac{2K}{\omega _0}}}~. 
$$

\bigskip

The fermionic zero modes can be obtained as the inverse of the bosonic ones. Thus
\begin{equation}\label{wf1}
{\rm w_1^+=\frac{1}{w_2^{+}(t;\alpha _+,\beta _+)}}\,,\quad {\rm w_1^-=\frac{1}{w_2^{-}(t;\alpha _-,\beta _-)}}~.
\end{equation}
A comparison of ${\rm w_1^+}$ with the common $1/\cos {\rm t}$ fermionic mode is displayed in Figures~(3) and (4).  

In the small ${\rm K}$ regime, ${\rm K} \ll \omega _0$, one gets
$$
{\rm w_2^{+}(t;\alpha _+,\beta _+) \approx\alpha _{+} z_{1}^{(p-\frac{1}{2})}
z_{2}^{(q-\frac{1}{2})}\, _{2}F_1\Big[2\, ,1\, , 2-\frac{K}{\omega _0} ; -\frac{1}{2}z_{1}(t)\Big]}-
$$
\begin{equation} \label{s1}
{\rm \beta  _+ 
e^{-2ip\pi}4^{(p-\frac{1}{2})}z_{1}^{-(p-\frac{1}{2})} z_{2}^{(q-\frac{1}{2})}\,   _2F_1\Big[\frac{K}{\omega _0}\, , 1+\frac{K}{\omega _0}\, ,\frac{K}{\omega _0}\, ; -\frac{1}{2}z_{1}(t)\Big]}
\end{equation}
and
$$
{\rm w_2^{-}(t;\alpha _-,\beta _-)\approx\alpha _{-}
z_{3}^{r} z_{4}^{s}\, _{2}F_1\Big[1, 2, 2+i\frac{K}{\omega _0}\,; \frac{1}{2}z_{3}(t)\Big]}+ 
$$
\begin{equation} \label{ss2}
{\rm \beta _{-}
4^r z_{3}^{-r} z_{4}^{s}\, _{2}F_1\Big[1-i\frac{K}{\omega _0},-i\frac{K}{\omega _0}, -i\frac{K}{\omega _0}\,; \frac{1}{2}z_{3}(t)\Big]}~.
\end{equation}

Examining the bosonic equations, one can immediately see that the resonant frequencies acquired resistive time-dependent losses whose relative strength is given 
by the parameter K. The fermionic equations having time-dependent real parts of the frequency can be interpreted as parametric oscillators which are also affected by losses
through the imaginary part. 

\medskip

\noindent
{\bf 5}. {\em More {\rm K} parameters.}

\noindent
A more general case in this scheme is to consider the following matrix Dirac-like equation
$$
\Bigg[\left( \begin{array}{cc}
0 & -{\rm i }\\
{\rm i} & 0\end{array} \right ){\rm D_{\rm t}}+\left( \begin{array}{cc}
0 & 1\\
1 & 0 \end{array} \right )\left( \begin{array}{cc}
 {\rm iu_p +K_1}& 0\\
0 &{\rm  iu_p+K_2}\end{array} \right )\Bigg]\left( \begin{array}{cc}
{\rm w}_1\\
{\rm w}_2 \end{array} \right )=
$$
\begin{equation} \label{Dg}
\left( \begin{array}{cc}
{\rm K_{1}^{'}}& 0\\
0 &{\rm  K_{2}^{'}}\end{array} \right )\left( \begin{array}{cc}
{\rm w_1}\\
{\rm w_2} \end{array} \right )~.
\end{equation}
The system of coupled first-order differential equations will be now
\begin{eqnarray}
\Big[-{\rm i}{\rm D_{\rm t}}+{\rm iu_p}+{\rm K_2}\Big]{\rm w_2}={\rm K_{1}^{'}}{\rm w_1}\\
\Big[{\rm i}{\rm D_{\rm t}}+{\rm iu_p}+{\rm K_1}\Big]{\rm w_1}={\rm K_{2}^{'}}{\rm w_2}
\end{eqnarray}
and the equivalent second-order differential equations
\begin{equation} \label{Schrgb}
{\rm D_{\rm t}}^{2}{\rm w} _{i}+\Big[-{\rm i}\Delta {\rm K}\Big]{\rm D_{\rm t}}{\rm w}_i
+\Big[\pm {\rm D_{\rm t}}{\rm u}_{\rm p}+{\rm i} ({\rm K}_1+{\rm K_2}){\rm u_p}+({\rm K_1K_2-K_{1}^{'}K_{2}^{'}})
 -{\rm u_{p}^2}\Big]{\rm w} _{i}=0~,
\end{equation}
where the subindex $i=1,2$ and $\Delta {\rm K}={\rm K _1}-{\rm K _2}$. 
Under the gauge transformation 
\begin{equation} \label{mg1}
{\rm w} _{i}={\rm Z}_{i}\exp \left(-\frac{1}{2}\int ^{\rm t}\Big[-{\rm i}\Delta {\rm K}\Big] d\tau\right)={\rm Z}_i({\rm t})e^{\frac{1}{2}{\rm i\, t} \Delta {\rm K}}~,
\end{equation}
one gets
\begin{equation} \label{schz}
{\rm D_{\rm t}}^{2}{\rm Z}_i+Q_i({\rm t}){\rm Z}_i=0,
\end{equation}
where the `potentials' have the form
\begin{equation} \label{Q}
Q_{i} ({\rm t})=\Big[\pm {\rm D_{\rm t}}{\rm u}_{\rm p}+{\rm i}({\rm K_1}+{\rm K_2}){\rm u_p}+({\rm K_1K_2-K_{1}^{'}K_{2}^{'}})
 -{\rm u_{p}}^{2}\Big]-
\frac{1}{4}\Big[-{\rm i}\Delta {\rm K}\Big]^2
\end{equation}
$Q_{1,2}$ are functions that differ from the nonoperatorial parts in Eqs.~(\ref{comp1} - \ref{comp2b}) only by constant terms.
Indeed, one can obtain easily the following equations:

For the fermionic spinor component one gets
\begin{equation} \label{comp1} 
{\rm D^{2}_{t}Z_1^{+}-\omega _0^2\Big[1+2\tan ^2 \omega _0 t-\frac{(K_2-K_1)^2}{4\omega _0^2} -\frac{K_1K_2-K_{1}^{'}K_{2}^{'}}{\omega _0^2}
+i\frac{K_1+K_2}{\omega _0}\tan \omega _0 t\Big] Z_1^{+}=0 }
\end{equation}
for $\kappa =1$, and 
\begin{equation} \label{comp1b} 
{\rm D^{2}_{t}Z_1^{-}+\omega _0^2\Big[1-2{\rm coth}^2 \omega _0 t +\frac{(K_2-K_1)^2}{4\omega _0^2} +\frac{K_1K_2-K_{1}^{'}K_{2}^{'}}{\omega _0^2}+i\frac{K_1+K_2}{\omega _0}{\rm coth} \,\omega _0 t\Big] Z_1^{-}=0 }
\end{equation}
for $\kappa =-1$. 

The bosonic component fulfills
\begin{equation} \label{comp2} 
{\rm D^{2}_{t}Z_2^{+}+\omega _0^2\Big[1 +\frac{(K_2-K_1)^2}{4\omega _0^2} +\frac{K_1K_2-K_{1}^{'}K_{2}^{'}}{\omega _0^2} -i\frac{K_1 +K_2}{\omega _0}\tan \omega _0t\Big] Z_2^{+}=0}~,
\end{equation}
 for $\kappa =1$, and 
\begin{equation} \label{comp2b} 
{\rm D^{2}_{t}Z_2^{-}-\omega _0^2\Big[1 -\frac{(K_2-K_1)^2}{4\omega _0^2} -\frac{K_1K_2-K_{1}^{'}K_{2}^{'}}{\omega _0^2} -i\frac{K_1 + K_2}{\omega _0}{\rm coth} \,\omega _0t\Big] Z_2^{-}=0}~,
\end{equation}
 for $\kappa =-1$. When ${\rm K_1=K_2=K}$ one gets the particular case studied in full detail above.

The more general `bosonic' modes have the form:
\begin{eqnarray}
{\rm Z_{2}^{+}(t;\alpha_{+},\beta_{+})=\alpha_{+}[\tan(\omega _0t) -{\rm i}]^{\frac{\Omega _1}{4\omega _0}}
[\tan(\omega _0t) +{\rm i}]^{\frac{\Omega _2}{4\omega _0}}}\, \nonumber\\
_{2} {\rm F}_{1}\left[\frac{\Omega _1 +\Omega _2}{4\omega _0},\frac{\Omega _1 +\Omega _2}{4\omega _0}+1\,,
1+\frac{\Omega _1}{2\omega _0};\frac{1}{2}(\tan (\omega _0 {\rm t})-{\rm i})\right]\nonumber\\
 {\rm +\beta_{+}(-1)^{-\frac{\Omega _1}{2\omega _0}}[\tan(\omega _0t) -{\rm i}]^{-\frac{\Omega _1}{4\omega _0}}[\tan(\omega _0t) +{\rm i}]^{\frac{\Omega _2}{4\omega _0}}}\,\nonumber\\
_{2} {\rm F}_{1}\left[\frac{\Omega _2 -\Omega _1}{4\omega _0},\frac{\Omega _2 -\Omega _1}{4\omega _0}+1\,,
1-\frac{\Omega _1}{2\omega _0};\frac{1}{2}(\tan (\omega _0 {\rm t})-{\rm i})\right]\nonumber\\
\end{eqnarray}
and
\begin{eqnarray}
{\rm Z}_{2}^{-}({\rm t};\alpha_{-},\beta_{-}) = 
\alpha_{-}[{\rm coth}(\omega _0{\rm t}) -1]^{\frac{\Omega _3}{4\omega _0}}
[\coth(\omega _0{\rm t}) +1]^{\frac{\Omega _4}{4\omega _0}}\, \nonumber\\
_{2} {\rm F}_{1}\left[\frac{\Omega _3 +\Omega _4}{4\omega _0}+1,\,\frac{\Omega _3 +\Omega _4}{4\omega _0},
1+\frac{\Omega _3}{2\omega _0};-\frac{1}{2}({\rm coth} (\omega _0 {\rm t})-1)\right]\nonumber\\
 +{\rm \beta_{-}(-1)^{-\frac{\Omega _3}{2\omega _0}}4^{\frac{\Omega _3}{2\omega _0}}[{\rm coth}(\omega _0{\rm t}) -1]^{-\frac{\Omega _3}{4\omega _0}}[{\rm coth}(\omega _0{\rm t}) +1]^{\frac{\Omega _4}{4\omega _0}}}\,\nonumber\\
_{2} {\rm F}_{1}\left[\frac{\Omega _3 -\Omega _4}{4\omega _0},\frac{\Omega _4 -\Omega _3}{4\omega _0}+1\,,
1-\frac{\Omega _3}{2\omega _0};-\frac{1}{2}({\rm coth} (\omega _0 {\rm t})-1)\right]~,\nonumber\\
\end{eqnarray}

where
$$
{\rm \Omega _1=
\left(4\omega _0^2+(K_1+K_2)^2+4[(K_1+K_2)\omega _0-K_{1}^{'}K_{2}^{'}]\right)^{1/2}}~,
$$
$$
{\rm \Omega _2=
\left(4\omega _0^2+(K_1+K_2)^2-4[(K_1+K_2)\omega _0+K_{1}^{'}K_{2}^{'}]\right)^{1/2}}~,
$$
$$
{\rm\Omega _3=
\left(4\omega _0^2-(K_1+K_2)^2-4[{\rm i}(K_1+K_2)\omega _0-K_{1}^{'}K_{2}^{'}]\right)^{1/2}}~,
$$
$$
{\rm\Omega _4=
\left(4\omega _0^2-(K_1+K_2)^2+4[{\rm i}(K_1+K_2)\omega _0+K_{1}^{'}K_{2}^{'}]\right)^{1/2}}~.
$$


\bigskip

\noindent
{\bf 6.} {\em Applications.}\\

{\em 6.1 Waveguides.}

\noindent
In view of the correspondence between mechanics and optics, one can also provide an interpretation in terms of the Helmholtz optics for light propagation 
in waveguides of special profiles. The supersymmetry of the Helmholtz equation has been studied by Wolf and collaborators \cite{W2}. 
To get the waveguide application, one should switch from the temporal independent variable to a spatial variable ${\rm t\rightarrow x}$ along which we consider
the inhomogeneity of the fiber whereas the propagation of beams is along another supplementary spatial coordinate $\rm z$.
Thus, we turn the equations (14-17) into Helmholtz waveguide equations of the type (we take $\rm c=1$)
 \begin{equation}\label{H1} 
{\rm 
\large[\partial _z^2 +\partial  _x^2+\omega _0 ^2n^2(x)\large]\varphi(x,z)=0}~,
\end{equation}  
where the modes $\varphi(x,z)$ can be written in the form $\rm w_{1,2}(x) e^{-ik_z z}$ for a fixed wavenumber $\rm k_z$ in the propagating coordinate that is common to
both wavefunctions and the index profiles correspond to two pairs of bosonic-fermionic waveguides and are given by
\begin{equation} \label{H2}
{\rm n_{b}^2(x) \sim 1- i\frac{2K}{k _0}\tan(k _0x)}~, \quad {\rm  n_{f}^2(x)\sim -(1+2\tan ^2(\omega _0 x))- i\frac{2K}{k _0}\tan(k _0x)}~,
\end{equation} 
and
\begin{equation} \label{H3}
 {\rm  n_b^2(x)\sim -1- i\frac{2K}{k _0}{\rm coth}(k _0x)~, \quad  n_f^2(x) \sim 1-2coth ^2(k _0 x)+ i\frac{2K}{k _0}{\rm coth}(k _0x)}~,
\end{equation}
respectively. In our units $\rm k_0 = \omega _0$.
Eqs.~(\ref{H2}, \ref{H3}) can be obtained from Riccati equations of the type ($\rm c\neq 1$)
\begin{equation}\label{H4}
\rm \omega _0^2n_{\rm f,b}^2(x)/c^2=k^2\mp R_x-R^2~,
\end{equation}
where $\rm R(x)$ are Riccati solutions directly related to the Riccati solutions discussed in the previous sections.

\medskip

According to Chumakov and Wolf \cite{W2} a second waveguide interpretation is possible describing two different Gaussian beams, 
bosonic and fermionic, whose small difference in frequency is given in terms of a small parameter $\epsilon$ (wavelength/beam width), 
propagating in the {\em same} waveguide. In this interpretation, the index profile is the same for both beams. For illustration,
let us take the normal oscillator Riccati solution in the space variable $\rm x$, i.e., $\rm tan k_0 x$ that we approximate to first order linear Taylor term $\rm  k_0 x$.
Then, the two beam interpretation leads to the following Riccati equation (for details, see the paper of Chumakov and Wolf)
\begin{equation} \label{CW1}
{\rm \omega _{1,2}^2n^2(x)-\omega _{0}^2n^2(0)=\mp k_0-k_0^2x^2(1\mp \epsilon)}~.
\end{equation} 
An almost exact, up to nonlinear corrections of order 
$\epsilon ^2$ and higher, supersymmetric pairing of the 
$\rm z$ wavenumbers (propagating constants) occurs, except for the `ground state' one. As noted by Chumakov and Wolf, supersymmetry connects in this case light
beams of different frequencies but having the same wavelength in the propagation direction $\rm z$. This approach is valid only in the paraxial approximation. Therefore,
one should know the small x behaviour of the K-modes in order to hope to detect them through stable interference patterns along the waveguide axis. 

\medskip

{\em 6.2 Cavity physics.}

\noindent
Another very interesting application of the K-modes in a radial variable could be Schumann's resonances, i.e., 
the resonant frequencies of the sperical cavity provided by the Earth's surface and the ionosphere plasma layer
\cite{CF04}. The Schumann problem can be approached as a spherical Helmholtz equation $[\nabla ^2_{\rm r} +k^2] \phi =0$ 
with Robin type (mixed) boundary condition $\frac{\partial \phi}{\partial n}|_{S}=C(\omega) \phi _S$, where $C(\omega)$ is expressed in terms of the skin depth
$\delta=\sqrt{2/(\mu _c \sigma \omega)}$ of the conducting wall, $\mu _c$ is its permeability and $\sigma$ is its conductivity. 
The eigenfrequencies fulfilling such boundary conditions can be written as follows
\begin{equation} \label{S1}
\omega ^2 \approx \omega _0 ^2 [(1-I) +{\rm i}I]~,
\end{equation}
where $I$ is a complicated expression in terms of skin depths and surface and volume integrals of Helmholtz solutions with 
Neumann boundary conditions
$\frac{\partial \phi}{\partial n}|_{S}=0$.
It is worth noting the similarity between these improved values of Schumann's eigenfrequencies and the K-eigenfrequencies. 
Moreover, using the $Q$ parameter of the cavity, one can write Eq.~(\ref{S1}) in the form
\begin{equation} \label{S2}
\omega ^2 \approx \omega _0 ^2 \Bigg[\left(1-\frac{1}{Q}\right) +{\rm i}\frac{1}{Q}\Bigg]~.
\end{equation}
This form shows that the modification of the real part of $\omega$ leads to a downward shift of the resonant frequencies, while the contribution to the imaginary 
component changes the rate of decay of the modes.

We point out that Jackson mentions in his textbook that the near equality of the real and imaginary parts of the change in $\omega ^2$ is a 
consequence of the employed boundary condition, which is appropriate for relatively good conductors.Thus, by changing 
the form of $C(\omega)$ that could result from different surface impedances, 
the relative magnitude of the real and imaginary parts of the
change in $\omega ^2$ can be made different. It is this latter case that corresponds better to the K-modes.  

\medskip

{\em 6.3 Crystal models.}

\noindent
There is also a strong mathematical similarity between the K-modes and the solutions of Scarf's crystal model \cite{Scarf58} based on the singular potential
$V(x)=-V_0{\rm cosec} ^2(\pi x/a)$, where $a$ is an arbitrary lattice parameter. For this model the one-dimensional Schr\"odinger equation has the form
\begin{equation}\label{scarf1}
\psi '' +(a/\pi)^2 \bigg[\lambda ^2 +\left(\frac{1}{4}-s^2\right){\rm cosec}^2(\pi x/a)\bigg]\psi =0~.
\end{equation}
For $0<x\leq a/2$, the general solution is
$$
\psi =[f(x)]^{\frac{1}{2}+s}\, _2{\rm F}_1\Big[\frac{1}{4}+\frac{1}{2}(s+\lambda), \frac{1}{4}+\frac{1}{2}(s-\lambda); 1+s; f ^2 (x)\Big]+
$$
\begin{equation}\label{scarf2}
 [f(x)]^{\frac{1}{2}-s}\, _2{\rm F}_1\Big[\frac{1}{4}-\frac{1}{2}(s-\lambda), \frac{1}{4}-\frac{1}{2}(s+\lambda); 1-s; f^2 (x)\Big]~,
\end{equation} 
where $f(x)=\sin (\pi x/a)$ corresponds to the ${\rm z_i}(t)$ functions, and $s$ and $\lambda$ corresponding to -p and -q, respectively, 
are related to the potential amplitude and energy spectral parameter. 
Thus, by turning the K-oscillator equations into corresponding Schroedinger equations, one could introduce another 
analytical crystal model with possible applications in photonics crystals.

\medskip

{\em 6.4 Cosmology.}

\noindent
Two of the authors applied the K-mode approach to barotropic FRW cosmologies \cite{RLS04}. K- Hubble cosmological parameters 
have been introduced and  expressed as
logarithmic derivatives of the K-modes with respect to the conformal time. For $\rm K \rightarrow 0$ the ordinary solutions of the common 
FRW barotropic fluids have been obtained.  

It is also worth noticing the analogy of the nonzero ${\rm K}$ oscillator case with the phenomenon of diffraction of atomic waves in 
imaginary crystals of light (crossed laser beams) \cite{crystL}.
In fact, the ${\rm K}$ parameter is a counterpart of the modulation parameter ${\rm Q}$ introduced by Berry and O'Dell in their study of 
imaginary optical gratings.
Roughly speaking, the nonzero ${\rm K}$ modes could ocurr in an {\em imaginary crystal of time} that could occur in some exotic astrophysical conditions. 

\bigskip

{\bf 7}. {\em Conclusion}.\\
By a procedure involving the factorization
connection between the Dirac-like equations and the simple second-order linear differential equations of harmonic oscillator type, a class of classical
modes with a Dirac-like parameters describing their damping and absorption (dissipation)  
has been introduced in this work.
While for zero values of the Dirac parameters the highly singular fermionic modes are decoupled from their normal bosonic harmonic modes,
at nonzero values a coupling between the two types of modes is introduced at the level of the matrix equation. These interesting modes are given
by the solutions of the Eqs.~(\ref{comp1}) - (\ref{comp2b}) and in a more general way by Eqs.~(\ref{mg1}, 34-35) of this work and are expressed 
in terms of hypergeometric functions.
Several possible applications in different fields of physics are mentioned as well. 
Finally, similar to the fact that the PT quantum mechanics can be considered as a complex extension of standard quantum mechanics, 
we notice that what we have done here is a particular type of complex extension of the classical harmonic oscillator.

\section*{Acknowledgment}

\noindent
The third author wishes to acknowledge  the support of CONACyT through Project J-41452 and Millennium Initiative W-8001.

\newpage


\begin{figure}
\begin{center}
\includegraphics{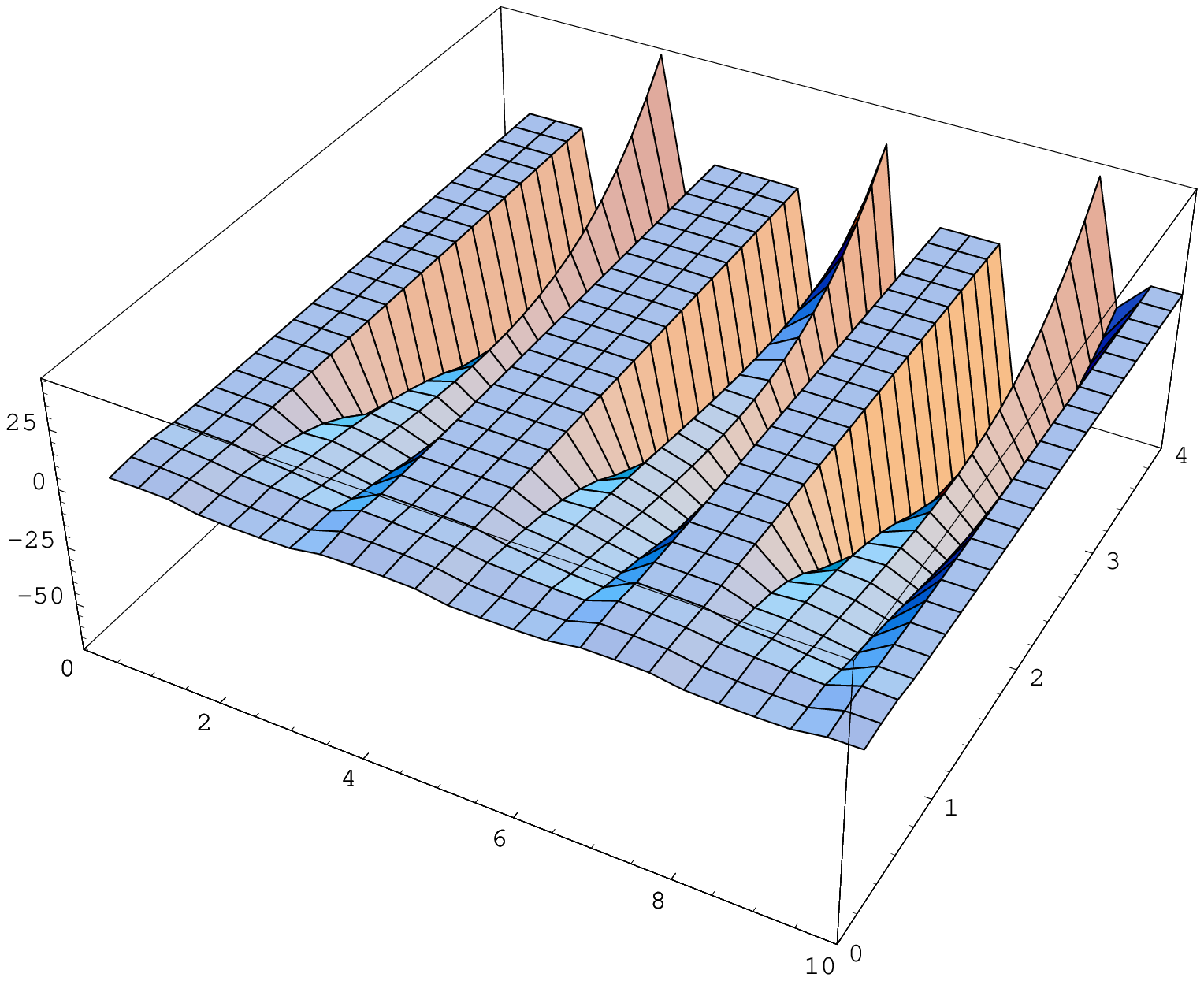}
\end{center}
\caption{The real part of the bosonic mode ${\rm w_2^{+}(y;\frac{1}{2},\frac{1}{2}})$ for ${\rm t}\in [0,10]$ and ${\rm K\in[0,4]}$.
}
\label{boring figure1}
\end{figure}


\begin{figure}
\begin{center}
\includegraphics{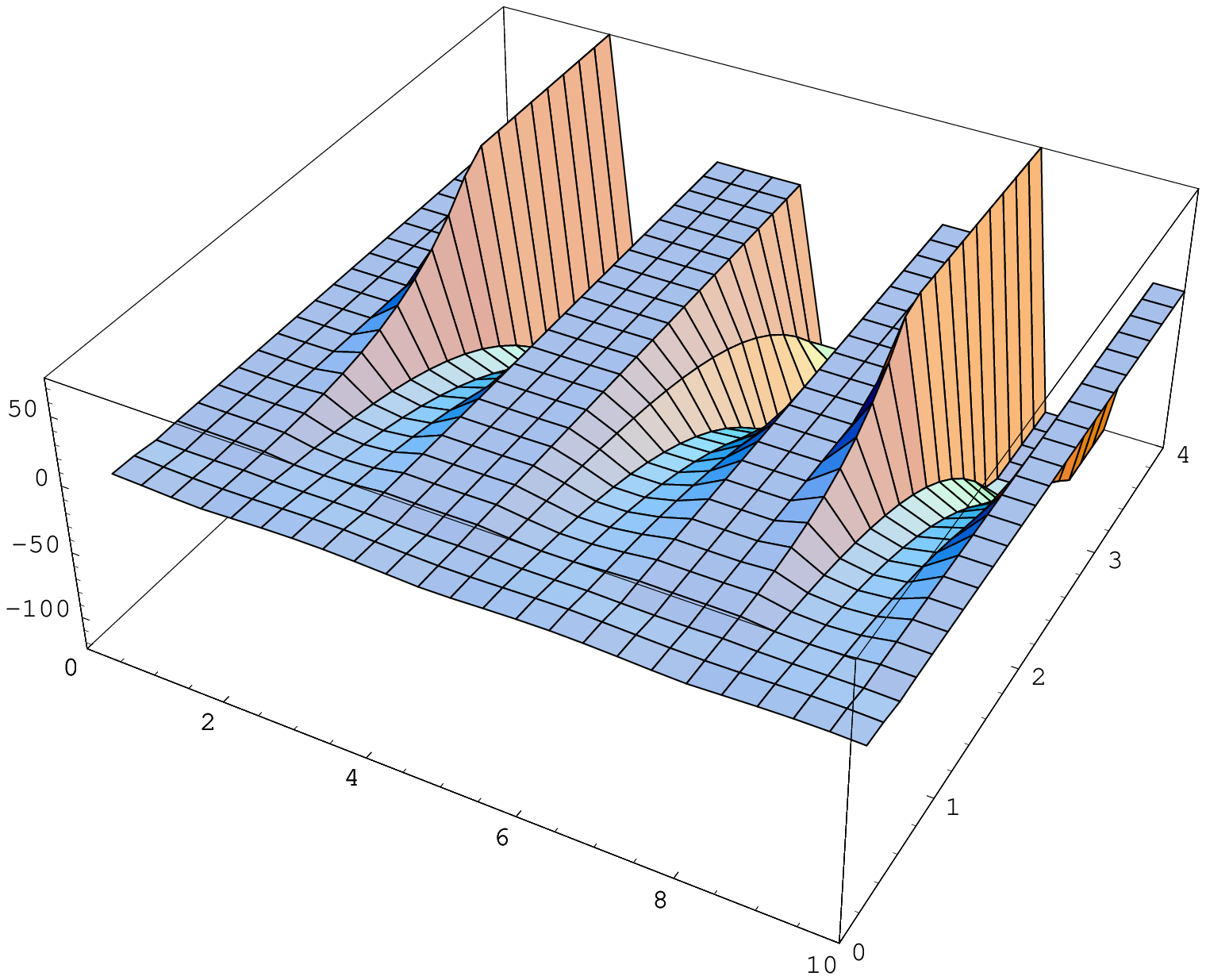}
\end{center}
\caption{The imaginary part of the bosonic mode ${\rm w_2^{+}(y;\frac{1}{2},\frac{1}{2}})$ for ${\rm t}\in [0,10]$ and ${\rm K\in[0,4]}$.}
\label{boring figure2}
\end{figure}


\begin{figure}
\begin{center}
\includegraphics{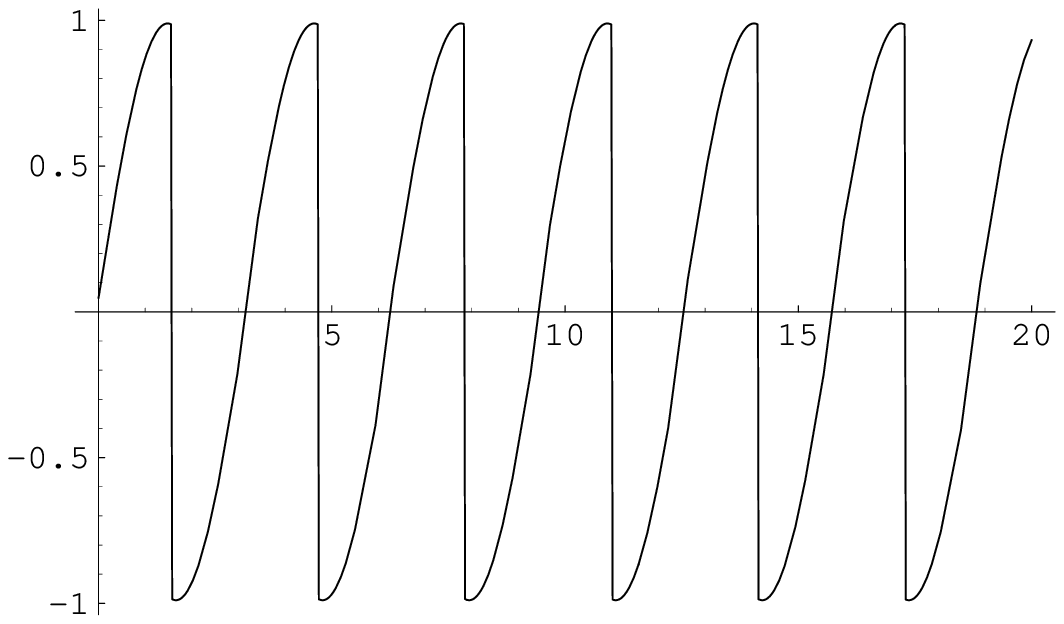}
\end{center}
\caption{The real part of the bosonic mode ${\rm w_2^{+}(y;\frac{1}{2},\frac{1}{2}})$ for ${\rm t}\in [0,20]$ and ${\rm K}=0.01$. 
}
\label{boring figure3}
\end{figure}


\begin{figure}
\begin{center}
\includegraphics{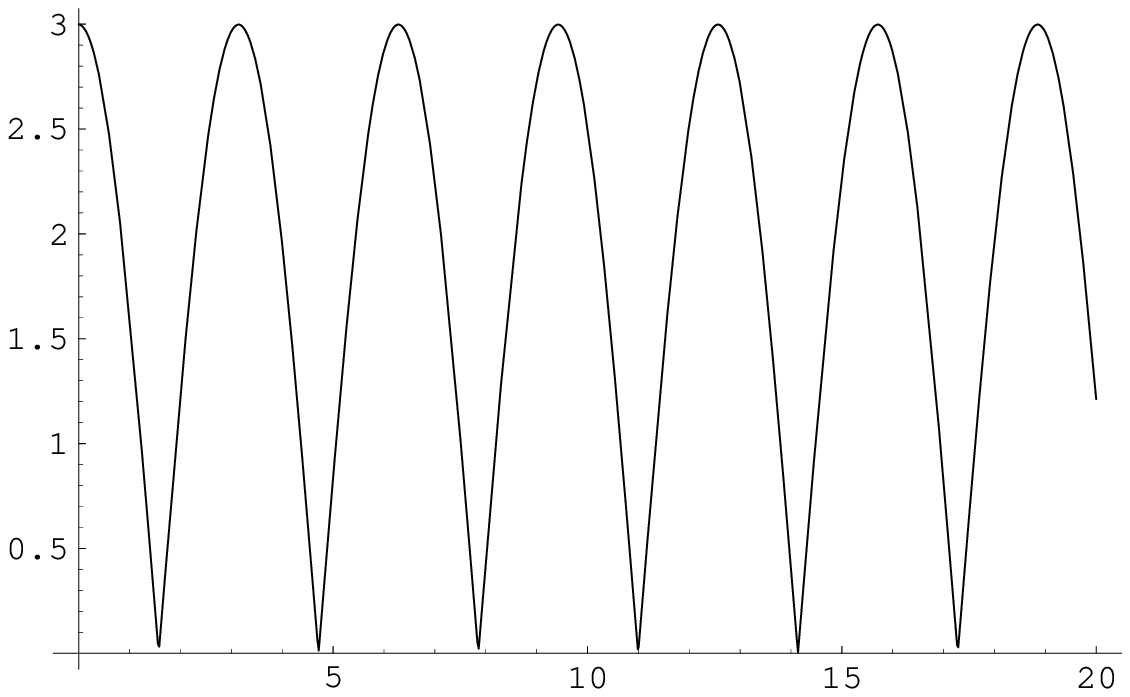}
\end{center}
\caption{
The imaginary part of the bosonic mode ${\rm w_2^{+}(y;\frac{1}{2},\frac{1}{2}})$ for ${\rm t}\in [0,20]$ and ${\rm K}=0.01$. 
}
\label{boring figure4}
\end{figure}


\begin{figure}
\begin{center}
\includegraphics{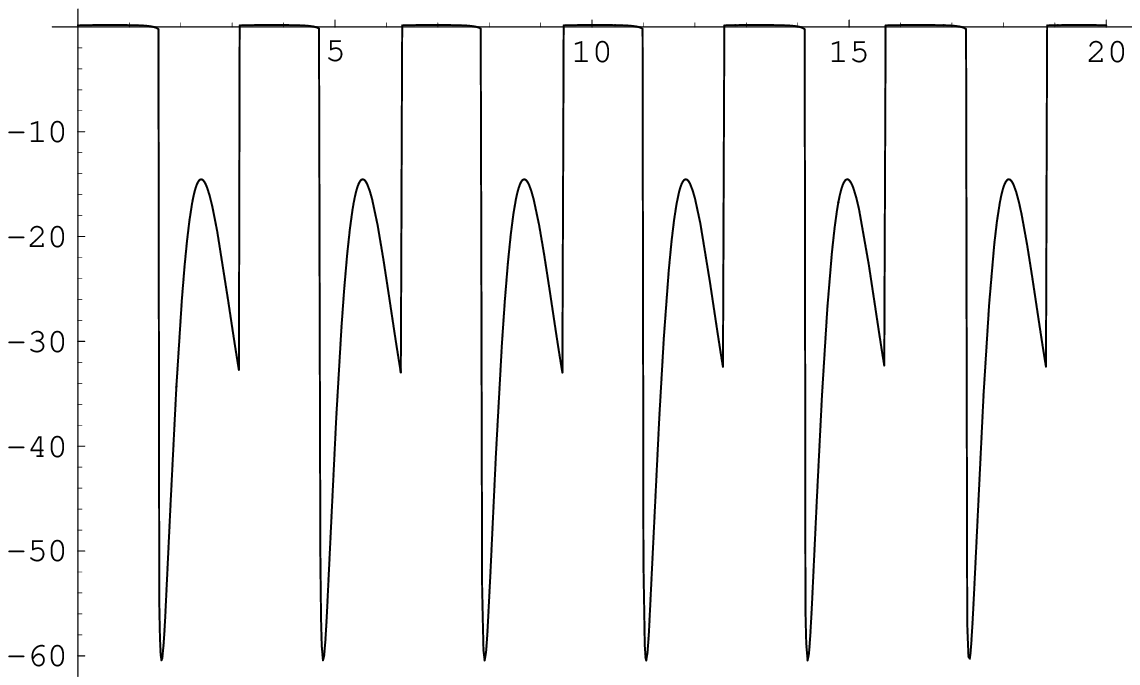}
\end{center}
\caption{The real part of the bosonic mode ${\rm w_2^{+}(y;\frac{1}{2},\frac{1}{2}})$ for ${\rm t}\in [0,20]$ and ${\rm K}=2$. }
\label{boring figure5}
\end{figure}


\begin{figure}
\begin{center}
\includegraphics{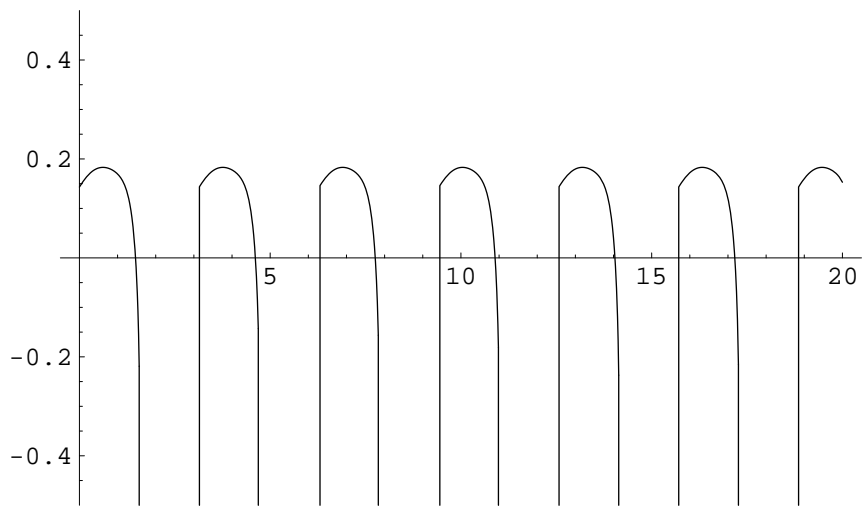}
\end{center}
\caption{The real part of the bosonic mode ${\rm w_2^{+}(y;\frac{1}{2},\frac{1}{2}})$ for ${\rm t}\in [0,20]$ and ${\rm K}=2$ in the vertical strip [-0.5, 0.5]. }
\label{boring figure5}
\end{figure}


\begin{figure}
\begin{center}
\includegraphics{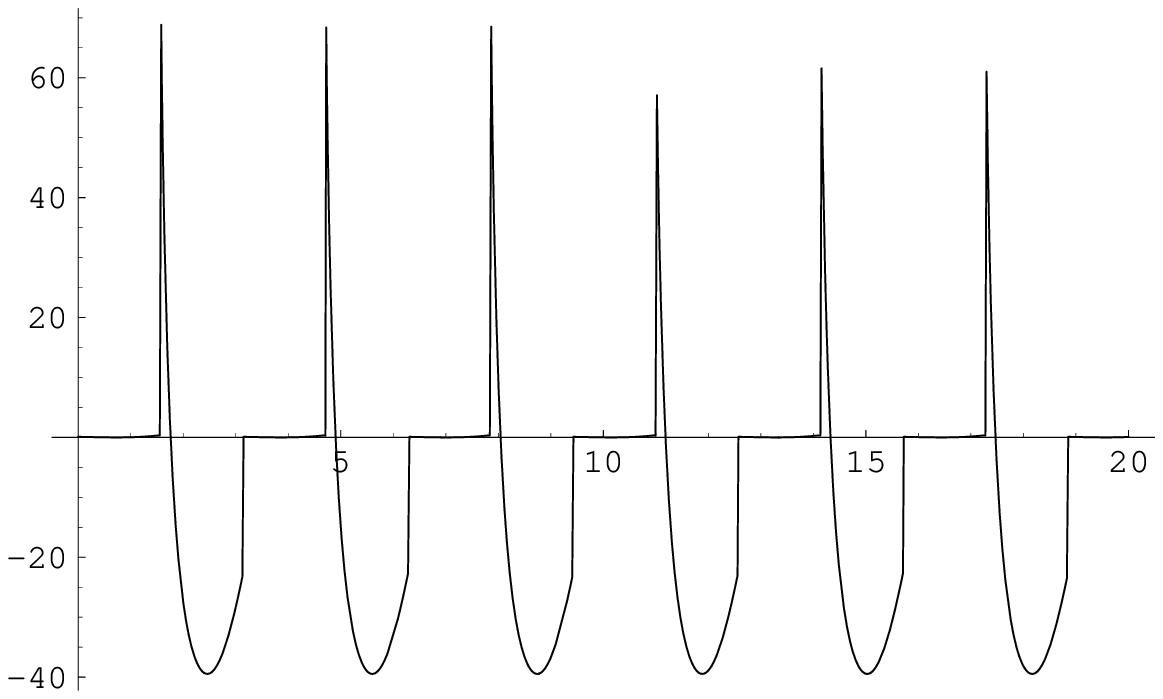}
\end{center}
\caption{
The imaginary part of the bosonic mode ${\rm w_2^{+}(y;\frac{1}{2},\frac{1}{2}})$ for ${\rm t}\in [0,20]$ and ${\rm K}=2$. 
}
\label{boring figure6}
\end{figure}


\begin{figure}
\begin{center}
\includegraphics{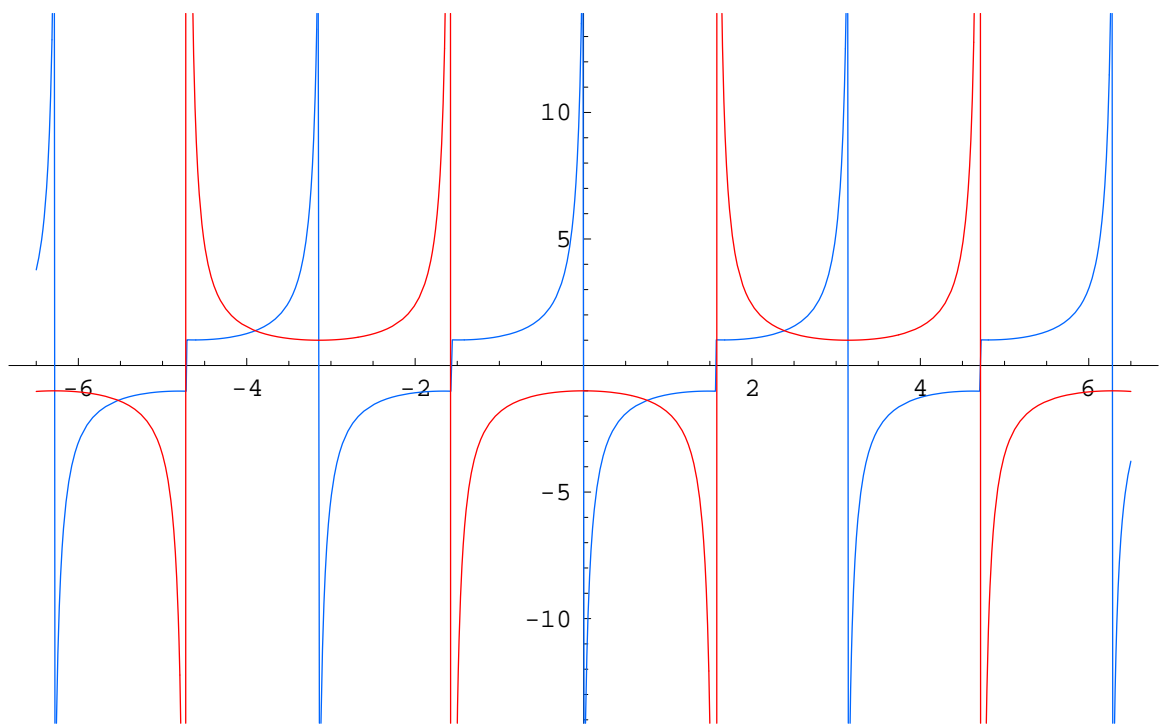}
\end{center}
\caption{
The fermionic zero mode $-1/\cos{\rm  t}$, (red curve), and the real part of $-1/{\rm w_2^{+}}$, (blue curve), for ${\rm K}=0.01$.}
\label{boring figure7}
\end{figure}


\begin{figure}
\begin{center}
\includegraphics{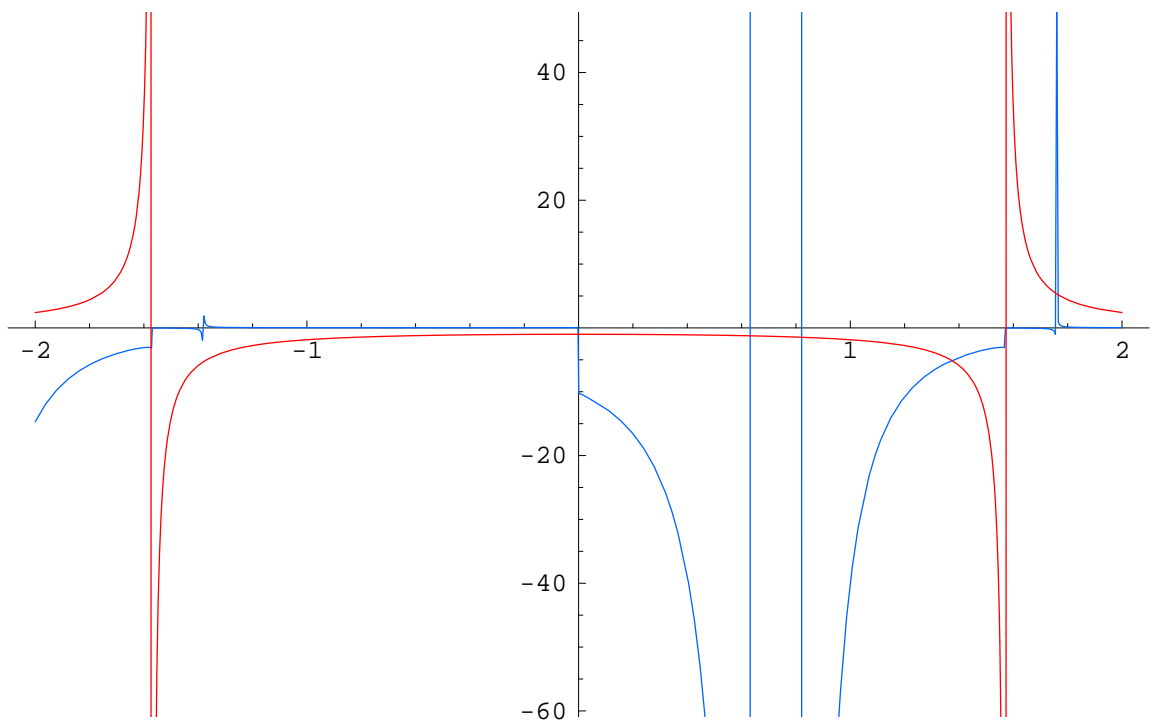}
\end{center}
\caption{
The fermionic zero mode $-1/\cos{\rm  t}$, (red curve), and the imaginary part of $-1/{\rm w_2^{+}}$, (blue curve), for ${\rm K}=2$.}
\label{boring figure8}
\end{figure}


\end{document}